# Superpositions between non linear intermittency maps. Application in biological neurons networks.

**Contoyiannis. F. Yiannis[1].**

1.  Department of Electrical and Electronics Engineering, University of West Attica, Ancient Olive Grove Campus, 250 Thivon and P. Ralli, Athens GR12244, Greece.

*Correspondence: yiaconto@uniwa.gr

Abstract:

In a series of works of ours we have shown that we can represent the critical and tricritical points of the Statistical Physics of critical phenomena as a Dynamical phenomenon expressed by time series produced by the type I intermittency that exhibits a weak chaos. Recently we have also shown that if we couple these two chaotic dynamics, namely critical and tricritical, we can produce a time sequence which is a temporal Spike Train (ST) of biological-type . In the present work we generalize this issue producing superpositions of critical-tricritical intermittencies with different parameter values. Now arise the question whether the coupling occurs between time series that have resulted from the superposition, will preserved or destroyed the ST biological type , as the number of intermittencies in the superposition will increase? In the other side in present work we find that the spikes produced by the chaotic dynamics of the intermittencies, under the action of superpositions and coupling remain biological-type too. Thus we can say that the dynamics of the fluctuations of the values of the time series produced by the coupling of the superpositions of the intermittencies is the same as the dynamics of the fluctuations of the membrane potential of the biological neuron. Given also that we can manipulate the numerical experiment of superposition and coupling as we wish, we will be able, in future, to approach the cause of neurological problems and decline in thinking ability due to loss of spikes in the brain.

Keysword: Intermittent Chaotic Dynamics, Critical phenomena in Physics, Spike trains in artificial neural networks and in Biological neural networks, Analysis timeseries.

### Section 1. INTRODUCTION

The phenomenon of intermittency ( in chaotic Dynamic) for a quantity Q(t) is the temporal alternation of regions of low fluctuations called laminar regions with chaotic regions which demonstrate high fluctuations ( bursts) . In figure 1 such a segment of a timeseries which present this alternation is shown.

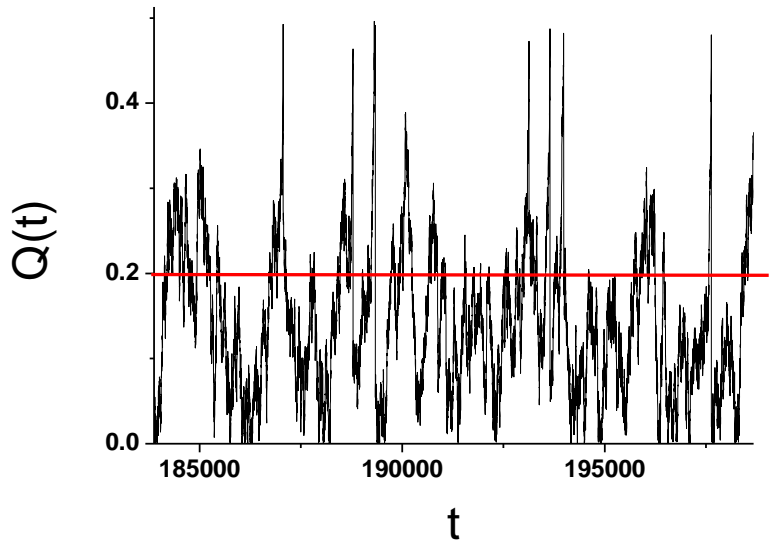

*Fig1. A segment of intermittent timeseries where we have suppose that the zone from 0 up to 0.2 ( red line ) is the laminar region and every value which is upper from this zone belong in bursts.*

The way to construct a numerical simulation through the intermittent map ( see below eq.1) to produce an intermittency time series as in figure 1 is to find its successive values through the following way : That is, as soon as an burst ends, the trajectory returning randomly to the values of the laminar region, e.g. [0,0.2] from fig.1 and so on. Considering the laminar region as the input values then the bursts are the output values . So, the return from output back to input of the laminar regions values has feedback character in this procedure is present .

The laminar lengths L are the waiting times inside the zone of laminar region of fig1 . Thus the laminar lengths determined through the relation:

$$\Phi_{red} < \Phi < \Phi_{blue}$$

The level $\Phi_{red}$ has a constant value which is the lower value of grass. The value value $\Phi_{blue}$ is a free parameter. So, a sweep of grass values inside this zone is accomplished. If we want investigate about a possible critical state of timeseries then the sweeps are finished when we find the best power-law distribution of laminar lengths.

It is known [1] that the Dynamic fluctuations of the order parameter Φ in the second order phase transition in critical state obey to the critical intermittency expressed mathematically with the intermittent map type I.

$$\Phi_{n+1} = \Phi_n + u_1 \Phi_n^z + \varepsilon_n \quad (1)$$

Where

$u_1 > 0$, $z > 0$ , $\varepsilon_n$ uniform noise between $[-\varepsilon_1 , +\varepsilon_1]$.

As we have shown in our work [1] the quantities that appear in equation 1 are quantities from physics. Thus as $\Phi_n$ we denote the fluctuations of the order

parameter and the exponent z is connected to the isotherm critical exponent δ with the relation

$$z = \delta + 1. \quad (2)$$

It is known from the theory of critical phenomena [2] that there is a region in parametric space where a crossover from the second-order phase transition to the first-order phase transition takes place. This crossover is accomplished around the so-called Griffiths tricritical point [2]. As we have shown in our work [3], the dynamics in the begging of this crossover is described by a mathematical intermittent map given as:

$$\Phi_{n+1} = \Phi_n - u_2 \Phi_n^{-z} + \varepsilon_n \quad (3)$$

Where $u_2 > 0$, $z > 0$, $\varepsilon_n$ uniform noise between [-$\varepsilon_2$, $\varepsilon_2$].

The negative sign of the non-linear term and the negative exponent ensures that the values fall. The two maps ( Eqs 1,3) are repellors where the fixed point in map 1 (eq.1) is the zero and lead the trajectory at higher values ( rise) while the fixed point in map2 (eq.3) is high value ,theoretical the infinity, and lead the trajectory at smaller values ( fall). The dynamics is determined by the distribution of the laminar lengths L of the intermittency. According to reference [4] for the

intermittency map this distribution is given by the power-law :

$$P(L) \sim L^{-p} \quad (4)$$

With p=z/(z-1). Thus, Due to eq. 2 we have that :

$$p = 1 + (1/\delta) \quad (5)$$

Given the fact that $1 < \delta < \infty$ from equation 5 results that the critical intermittency (eq1) exist for exponent p∈ [1,2). According to reference [4] for the tricritical map (eq3) this distribution is given by the eq. 4 where now p=z/(z+ 1). Due to eq. 2 we have that in tricritical dynamics

$$p = (\delta+1)/(\delta+2) \quad (6).$$

Given the fact that $1 < \delta < \infty$ from equation 6 results that the tricritical intermittency exist for exponent p∈[0.66,1). All the above are in the framework of the Method of critical fluctuations (MCF )[5] with which we reveal how close the distribution of laminar lengths (generally waiting times) is to a power law. Now, If we find an exponent p ∈[1,2) then the system which produce the time series is at critical state too. In the other side if we find a exponent p∈ [0.66,1) then the system which produce this time series is at tricritical state. The fitting function which we use in order to calculate the exponent of power-law from the distribution of laminar lengths is :

$$f(L) \sim L^{-p_2} \cdot e^{-p_3 L} \quad (7)$$

As we see we have introduce in eq.7 an exponential corrective term where the exponent $p_3$ appears. When the $p_3$ is close to zero the distribution reach to the scaling form of eq.4. Thus the exponent $p_3$ is a "measure" how close we are in criticality or tricriticality in each case if of course the corresponding conditions for $p_2, p_3$ are valid. The first scope of present work is to investigate if the criticality or the tricriticality remain when we accomplished superpositions for each map case every time where we change the parameter values in eqs (1,3).

In our work [6] we had shown that a coupling mechanism between a critical map and a tricritical map can to create structures of spikes train [7-10] of biological type. Thus, in the present work we investigate the main question : Remain this interesting property of coupling i.e the spikes train production, after the operation of superpositions in each map or this property is destroyed?

## Section 2. The superposition of maps

. Let us assume that we have k critical intermittent maps . What would then be their superposition? Now we will have an equation of the following form:

$$\sum_{i=1}^{k} x_{n+1,i} = \sum_{i=1}^{k} x_{n,i} + \sum_{i=1}^{k} u_i x_{n,i}^{z_i} + \sum_{i=1}^{k} \varepsilon_{n,i} \quad (8)$$

Where $u_i>0$ and $z_i>1$. The above sum ( eq8) is repeated for each n inside the loop n=1, N with N the number of iterations. It is interesting question to investigate whether the superposition (8) without being the critical map (eq 1) , due to the non-linear polynomial term it contains , could numerically produce a timeseries where demonstrate a power law distribution of the laminar lengths inside limits of criticality.

As we can see from eq. 8, each of the component maps has its own parameter values. The term that has the greatest impact on the dynamics of the map is the nonlinear term with parameters u, z because this term define the limits of laminar regions. Thus we will focus on the parameters of this term.

In the following of this section we will present results from equation 8.

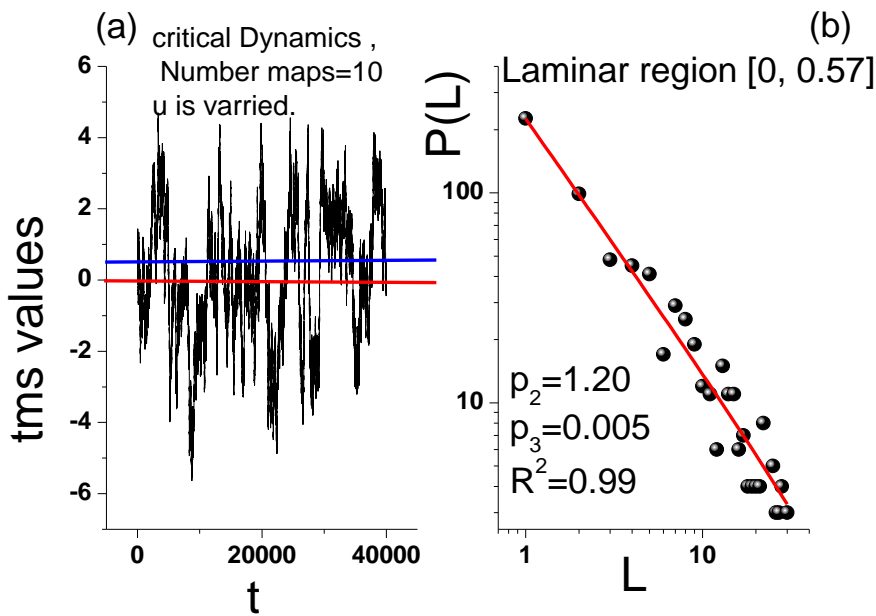

*Fig.2 (a) The time-series which produced for 40000 iterations from superpositions of 10 critical maps as the parameter u is varied with random way in the interval [0.02, 0.03]. The red and blue lines determined the laminar region [0, 0.57] where 0 is the fixed point of critical maps and 0.57 a free parameter. (b) The distribution of laminar lengths inside the laminar region. The exponent $p_2=1.20 \in$ [1,2], $p_3=0.005$ (very close to 0) . These values indicate that the distribution is a power-law in critical state.*

Let us now consider the case where the component maps do not have the same exponent z but different ones. As we have shown [11] , the exponent z is related to the firing rate, since e.g. a smaller value of the exponent means shorter stay times (laminar lengths) and therefore a greater firing rate. This is also the most realistic case, i.e. the components to have different firing rates.

The figure 3 shows the analysis for a superposition of 10 critical maps where all have the common coupling parameter u =0.011 , the width of noise ε in the range [-0.017, 0.017] but the exponents z are different in each neuron varying randomly in the interval : **integer** [6,2] namely z=2,3,4,5,6. The time series has a length of N=100000 . Here we find as laminar region the interval [0, 0.75] between the red and blue lines in fig. 3a and the corresponding laminar lengths distribution ( fig 3b) has exponent *p2=1.26 , p3=0.005.* These values indicate that the distribution is a power-law in critical state.

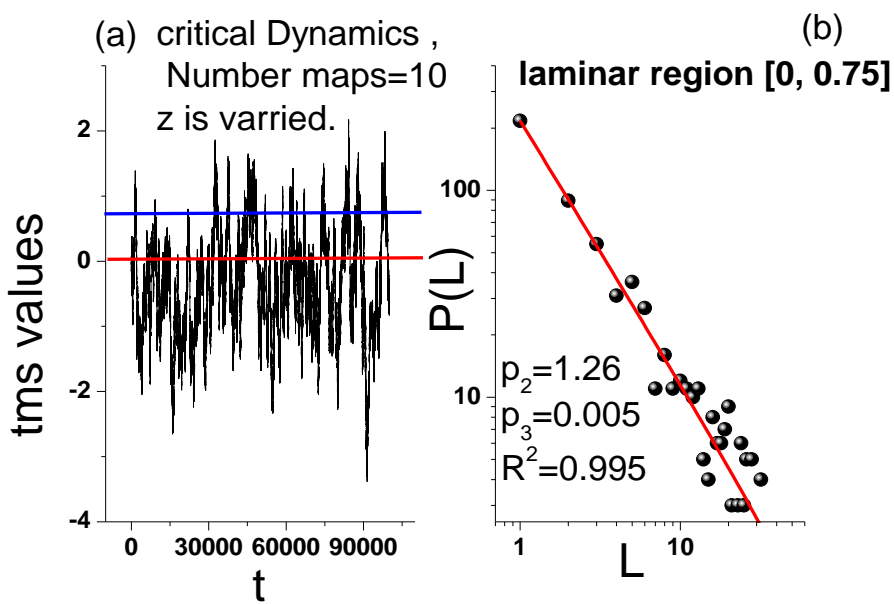

**Fig.3 *(a) The time-series which produced for 100000 iterations from superpositions of 10 critical maps as the parameter z is varied with random way in the interval : integer [6, 2] i.e 2,3,4,5,6. The red and blue lines determined the laminar region [0, 0.75]. (b) The distribution of laminar lengths inside the laminar region. The exponent p2=1.26∈ [1,2) , p3=0.005 . These values indicate that the distribution is a power-law in critical state.***

Let us assume now that we have k tricritical intermittencies. What would then be their superposition? We now will have an equation of the following form:

$$\sum_{i=1}^{k} x_{n+1,i} = \sum_{i=1}^{k} x_{n,i} - \sum_{i=1}^{k} u_i x_{n,i}^{-z_i} + \sum_{i=1}^{k} \varepsilon_{n,i} \quad (9)$$

Where the above sum ( eq.9) is repeated for each n inside the loop n=1, N with N the number of iterations and $u_i > 0$, $z_i > 1$.

The figure 4 shows the analysis for a superposition of 10 tricritical maps where all have the common coupling parameter u =0.2 , the width of noise ε in the range [-0.01, 0.01] but the exponents z are different in each map varying randomly in the interval **integer** [6,2] namely z=2,3,4,5,6. The time series has a length of N=100000 . Here we find as laminar region the interval [25,16] between the red and blue lines in fig. 4a and the corresponding laminar lengths distribution ( fig 4b) has exponent $p_2$=0.975 , $p_3$=0.005 . These values indicate that the distribution is a power-law in tricritical state.

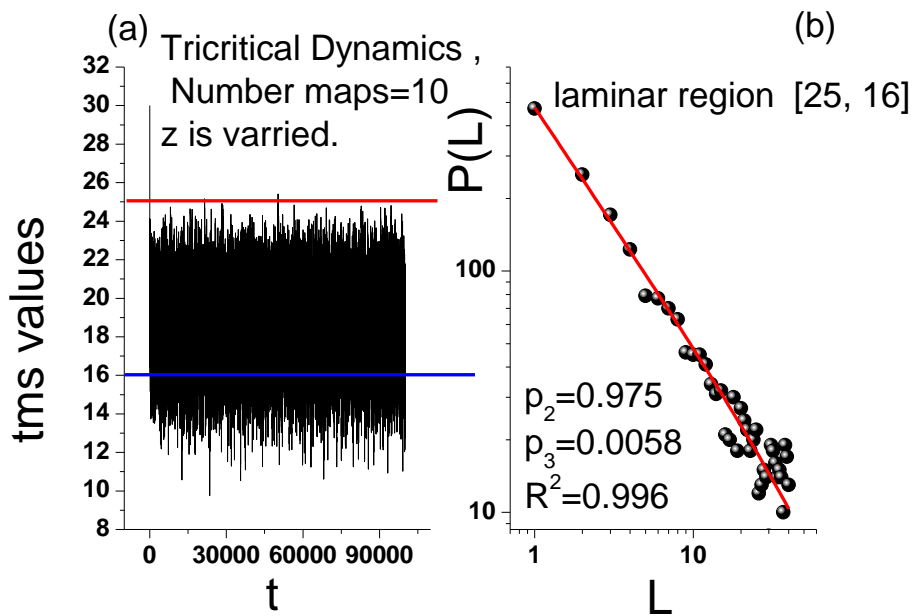

*Fig 4.(a) The time-series which produced for 100000 iterations from superpositions of 10 tricritical maps as the parameter z is varied with random way in the interval : integer [6, 2] i.e 2,3,4,5,6. The red and blue lines determined the laminar region [25, 16]. Here the great value of tricritical map is the fixed point because the values decrease (b) The distribution of laminar lengths inside the laminar region. The exponent $p_2=0.975 \in [0.66, 1)$, $p_3=0.005$ . These values indicate that the distribution is a power-law in tricritical state. The fixed point of each component is set to the value 3. So for 10 components the initial value is 30. Thus the red line should be at the value 30. However, between 30 and the line that we have set at 25, as can be seen from Figure 4, there is essentially no point, therefore the results do not change if we set the red line at 25.*

We will now proceed to superposition more maps, e.g. 100 maps.

The figure 5 shows the analysis for a superposition of 100 critical maps where all have the common coupling parameter u =0.2 , the width of noise ε in the range [-0.017, 0.017] but the exponents z are different in each neuron varying randomly in the interval **integer** [6,2] namely z=2,3,4,5,6. The time series has a length of N=100000. Here we find as laminar region the interval [2.5, 9] between the red and blue lines in fig. 5a and the corresponding laminar lengths distribution ( fig 5b) has exponent $p_2$=1.15 , $p_3$=0.016. These values indicate that the distribution is a power-law in critical state.

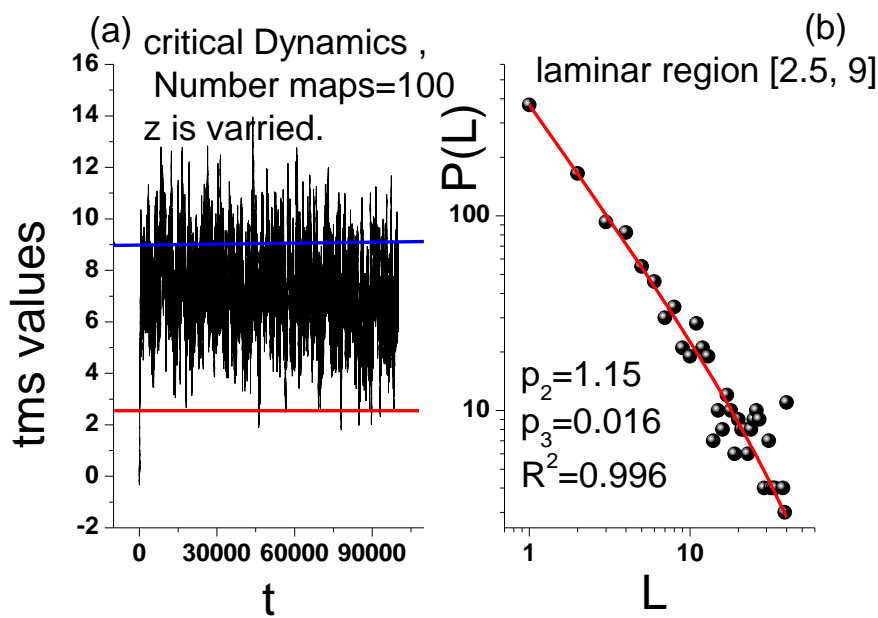

Fig.5 *(a) The time-series which produced for 100000 iterations from superpositions of 100 critical maps as the parameter z is varied with random way in the interval int [6, 2] i.e 2,3,4,5,6. The red and blue lines determined the laminar region [2.5, 9]. (b) The distribution of laminar lengths inside the laminar region. The exponent p2=1.15$\in$ [1,2) , p3=0.016 . These values indicate that the distribution is a power-law in critical state.*

The figure 6 shows the analysis for a superposition of 100 tricritical maps where all have the common coupling parameter u =1 , the width of noise ε in the range [-0.01, 0.01] but the exponents z are different in each map varying randomly in the interval: **integer** [6,2] namely z=2,3,4,5,6. The time series has a length of N=100000 . Here we find as laminar region the interval [200,168] between the red and blue lines in fig. 6a and the corresponding laminar lengths distribution ( fig 6b) has exponent $p_2$=0.77 , $p_3$=0.009 . These values indicate that the distribution is a power-law in the tricritical state

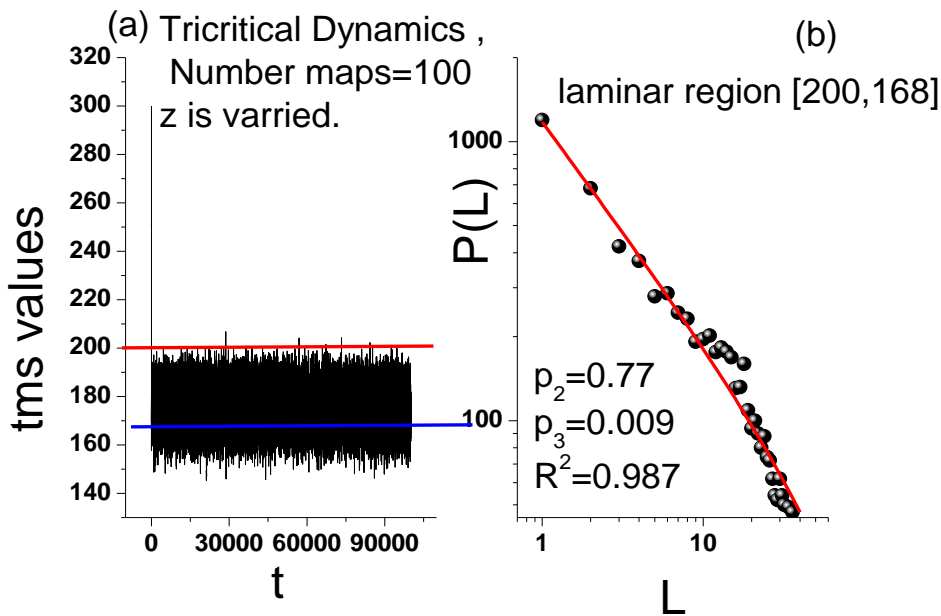

*Fig 6.(a) The time-series which produced for 100000 iterations from superpositions of 100 tricritical maps as the parameter z is varied with random way in the interval integer [6, 2] i.e 2,3,4,5,6. The red and blue lines determined the laminar region [200, 168]. (b) The distribution of laminar lengths inside the laminar region. The exponent $p_2$=0.77 $\in [0.66, 1)$ , $p_3$=0.009 . These values indicate that the distribution is a power-law in tricritical state. The fixed point of each component is set to the value 3. So for 100 components the initial value is 300. Thus the red line should be at the value 300. However, between 300 and the line that we have set at 200, as can be*

## Section 3.  The coupling mechanism between critical-tricritical fields.

In the previous section we showed that the two superpositions of the intermittencies maps (eqs. 1,3) do indeed have critical or tricritical dynamics. However, the mathematical formulas of the two superpositions (eqs. 8,9) while having the same dynamics as the critical and tricritical Dynamics (eqs. 1,3) cannot be characterized as critical or tricritical maps but as mathematical superpositions of them. Thus, we treat them as two time series that have arisen from eqs. 8, 9 . Therefore, the coupling mechanism that we used in our work [6] and concerned the coupling of the maps eqs. 1,3 must be modified. The basic change we make is for the case of time series (numerical or real). To achieve the drop in values in the time series that has the tricritical dynamic and has resulted from equation 9, the negative sign in the nonlinear polynomial term is not enough because the change of the parameters in the superposition we do not know what effect it will bring. The sure way is to strengthen the negative sign by changing the sign of the time series values of the tricriticality. Therefore, the general scheme of the coupling for the time series $X(i)$, i=1,N and $Y(i)$ i=1,N is:

**Coupling of laminar regions : X(i) ⇋ -Y(i).**

In following we present the code of coupling for the case of superpositions for 10 maps ( figs 3a, 4a).

The code for this coupling ( in Fortran) is the following:

```fortran
      dimension x(500000)
      dimension y(500000)
      dimension tms(500000)
      dimension coupltms(500000)
c     **************************************
      open(9,file="coupltms.dat")
      open(13,file="segment.dat")
      open(7,file="lamin100.dat")
c     ********Introduction data of critical  superposition ********
      open(12,file="C:\contoyiannis\Superposition\coupling\tmscr.for")
            do i=1,100000
                  read(12,*) x(i)
                  write(10,20)x(i)
            enddo
            close(12)
c     ********Introduction data of tricritical superposition *******
      open(14,file="C:\contoyiannis\Superposition\coupling\tmstrcr.for")

      do j=1,100000
            read(14,*) y(j)
                  write(11,20)y(j)
      enddo
      close(14)
c     *********Laminar region of  critical tms fig.1a*************
```

```fortran
      do i=1,100000
               if(x(i).lt.0.75.and.x(i).gt.0)then
                  tms(i)=x(i)
                  endif
       enddo
c      *********Laminar region of  - tricritical tms*************8
       do  j=1,100000
               if(y(j).lt.25.and. y(j).ge.16)then
                      k=1
                       tms(j+k)=-y(j)
                  endif
         enddo
       do l=1,100000
                coupltms(l)=tms(l)
           write(9,40)l,coupltms(l)

         enddo
c       ************segment*********
        do i= 19231  ,19921

            write(13,40)i,coupltms(i)
          enddo
     lamin=0

      do i=1,100000
       if (coupltms(i).lt.-14.and.coupltms(i).gt.-25)then
```

```
        lamin=lamin+1
       else
        write(7,10)lamin
        lamin=0
       endif
      enddo
 10   format(i8)
 20   format(f14.8)
 30   format(f14.8,5x,f14.8)
 40   format(i8,5x,f14.8)
      close(9)
      close(13)
     stop
     end
```

In Fig 7 we present the results of the coupling between the timeseries (fig 5a) and timeseries (fig.6a) i.e the case for 10 maps.

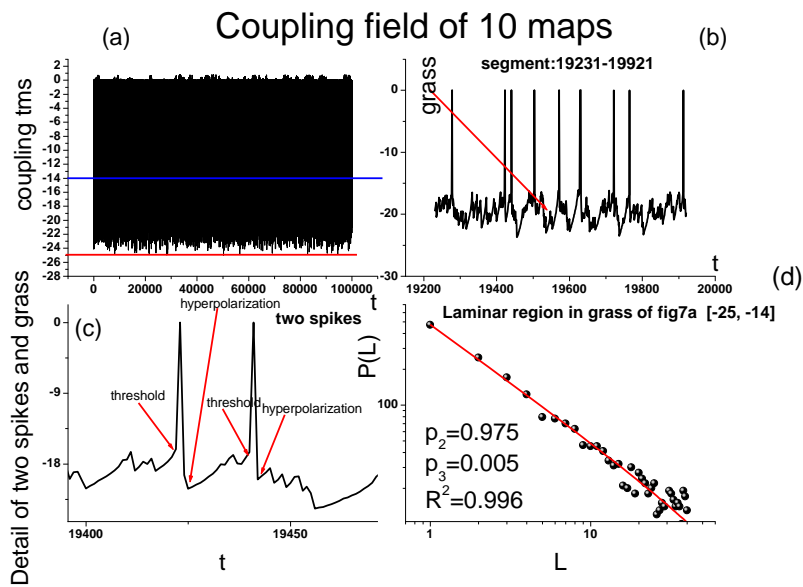

**Fig.7 (a) The timeseries of coupling for the case of 10 maps. Between the red and blue lines is the laminar region. (b) A segment from fig7a. (c) Two spikes from the fig 7b which are biological - type d) The laminar distribution inside the laminar region fig7a .** *These values indicate that the distribution is a power-law in tricritical state*

In Fig 8 we present the results of the coupling between the timeseries (fig 4a) and timeseries (fig.5a) i.e for 100 maps.

## Coupling field of 100 maps

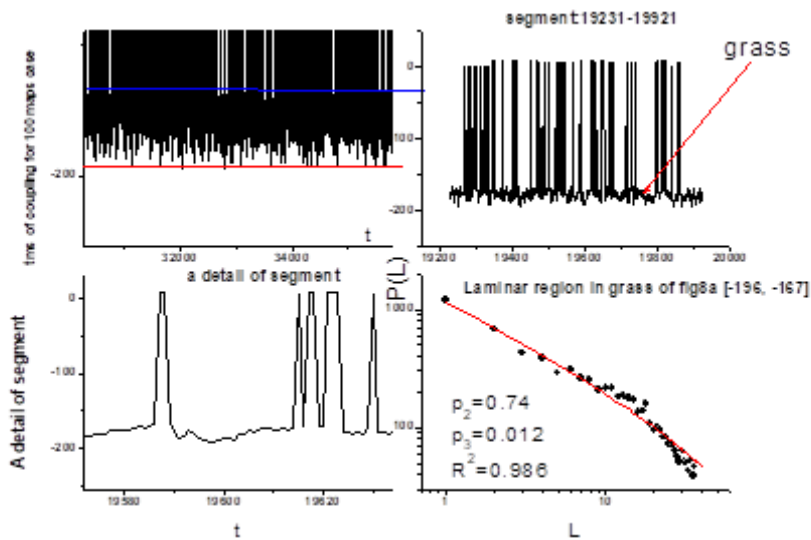

Fig.8 (a) The timeseries of coupling for the case of 100 maps. Between the red and blue lines is the laminar region. (b) A segment from fig 8a. (c) A detail of segment 8b  d) The laminar distribution inside the laminar region fig8a . These values indicate that the distribution is a power-law in tricritical state. In fig 8c some spikes, as we explain in following, have vanished and in its positions there are orthogonal or are structures connected between them.

As we see in fig8d the coupling of fields give in the relaxation periods (grass) between the temporal spikes tricritical Dynamics. This result do not exclude that changing the parameter values of maps the Dynamic in relaxation period could be a critical dynamics .

## Section 4. The Dynamic of the temporal Spike train for Biological neurons

As we refer in section 3 the main characteristics of Biological spin train are: the excitation threshold in rise period , the phenomenon of hyperpolarization in fall period  and the phenomenon of critical fluctuations in the grass zone in relaxation period.  In  Fig 9 we present  a timeseries  of  biological spikes .

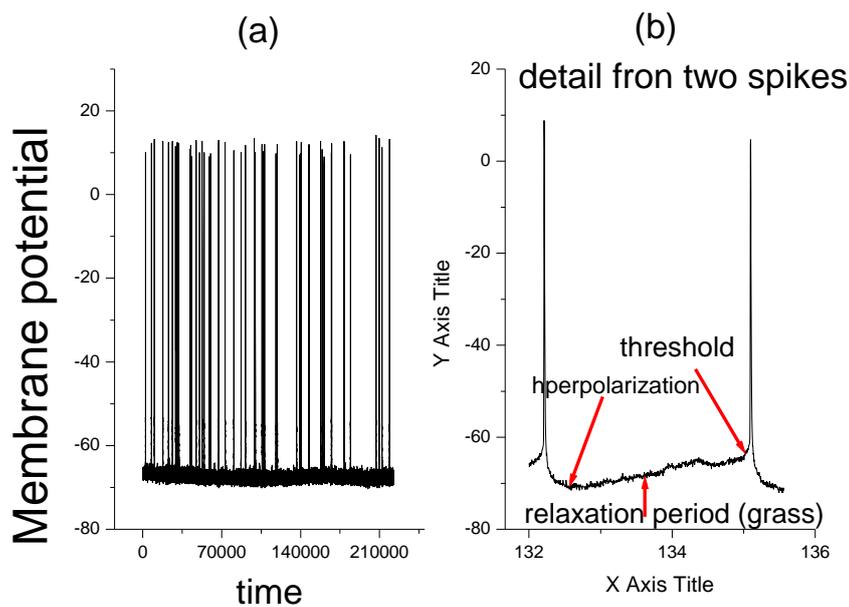

*Fig9.  (a)  A temporal  spike train  from biological membrane potential from in vitro intracellular recordings of CA1 pyramidal neurons in Wistar male rats [11] (b)  A detail  of two  spikes from fig9(a)  where the 3 main characteristic are shown. The similarity of the spike structure with the corresponding plot in 7(c) for the case of the fields with 10 maps is obvious.*

The biological data of the spike trains were obtained with an electrode that records the membrane potential from α biological neuron which communicates through synapses with at least 10000 neurons . In the case of the fields  of maps in our model  the corresponding is the fields of biological neurons. The our model based on two basic characteristic: The superposition and the coupling mechanism. Thus, the question now is : These operations exist in biological neurons ?

(a) What happens when the electrode records the membrane potentials from a field of neurons, and not from just one neuron. In this case, we have that the electrode records the superposition of the neurons of the area because it receives signals from all the neurons of the area simultaneously. This is consistency with the simultaneous character of superpositions of our model.

(b) The coupling in our model is characterized from the intertwined the of two maps as we seen in code (section 3). In our work [6] where we had produced type biological spikes from the coupling of the maps we had given the interpretation that for the creation of each spike peak the critical intermittency describes the dynamic of the Na ion pump i.e. raising the curve above the threshold value, while the tricritical maps describes the dynamic of the K ion pump (falling the curve below the threshold value.) Therefore the coupling phenomenon is appears also in the creation of biological spikes. Here the coupling is accomplished between the two biological pump ions of Na and ions of K.

Thus in two different systems i.e intermittency and biological neural network we have found the two assumptions i.e superpositions and coupling . Something like this allow us to characterize the two maps (1,3) with the word "neurons".

## Discussion-conclusions.

To answer the question we asked earlier, namely, whether by increasing the number of maps (or neurons) in a field the phenomenon of spike train creation remains or disappears, we compare the spike train in the 10 and 100 map superpositions (Figs. 7 c and 8 c). For reasons of normalized comparison, we have taken the time intervals of the segments at 10 and 100 map to be the same. Then we find that in the case of 10 maps in the segment interval there are 9 structures of which 8 are spikes, so the percentage of spikes is 8/9 X 100% = 88%. In the case of 100 maps we have 39 structures of which 17 are spikes and 22 are not, so the percentage is 17/39 X 100% = 44%. Therefore, the greater the number of maps- neurons in the superposition, the fewer spikes will be produced in the coupling. Thus, from the present work an important conclusion emerges for the biology of the brain. As the number of neurons in the field increases, the coupling between criticality and tricriticality (excitatory-inhibitory neurons) gives results that are far removed from the structure of the biological spike, since the edge of the spike is converted into a plane as in fig. 8c. This happens because the increase in the number of spikes causes overlap between them and thus they cease to be spikes. In other words, the number of spikes produced in the spike-train time sequence decreases. Something like this would justify neurological diseases due to the smaller number of spikes that flow in the neural networks of the brain, so the ability to transfer information and therefore thought will be reduced. On the other hand, if we want to keep the number of spikes constant, the time intervals of the spike train should be longer in the case of most neurons, i.e. the information that flows in the brain networks should be slower, something that also creates neurological problems. In this work we saw an example of superposition and coupling. However, the parameters of the problem are many and their combinations are multiples of them. Therefore, the variety of the spikes train produced is unmeasurable, something that justifies the complexity of the mechanism both in the numerical experiment with the dynamics of intermittencies and in the biological brain networks. Beyond the similarity in the chaotic intermittent dynamics, these are completely different systems. We have seen such systems in Nature or in technology [12] so far in earthquakes, and in hybrid artificial neurons networks.

For biological neural networks, we would propose the synapses and especially the electrical type synapses as the place where the superposition of neurons and their coupling occurs. This proposal is consistent with a series of works. Indeed, in the case of electrical synapses, (a) it is known [13-15] that neurons are bidirectionally coupled to each other through gap junctions[16]. (b) These types of synapses are also known [17] to produce synchronous network activity in the brain i.e. a superposition phenomenon occurs. This certainly does not exclude that electrical synapses can also lead to complex, chaotic dynamics at the network level.[18], [19].

# Bibliography

1. Contoyiannis, Y.F., Diakonos, F.K., Malakis. Intermittent Dynamics of Critical Fluctuations. Physical Review Letters, 2002, 89(3).

2. Huang K 1987 Statistical Mechanics 2nd (New York: Wiley).

3. Contoyiannis, Y., Potirakis, S.M., Eftaxias, K., Contoyianni, L. Tricritical crossover in analyzing preseismic electromagnetic emissions. Journal of Geodynamics, 2015, 84, pp. 40–54.

4. H. G. Schuster, Deterministic Chaos (VCH, Weinheim, 1998).

5. Yiannis Contoyiannis and Stelios M Potirakis J. Signatures of the symmetry breaking phenomenon in pre-seismic electromagnetic emissions. Stat. Mech. (2018) 083208.

6. Stelios M Potirakis, Fotios K Diakonos, Yiannis F Contoyiannis. A Spike Train Production Mechanism Based on Intermittency Dynamics. Entropy, 27(3), 2025.

7. Maass W (1997). "Networks of spiking neurons: The third generation of neural network models". Neural Networks. 10 (9): 1659–1671. doi:10.1016/S0893-6080(97)00011-7. ISSN 0893-6080.

8. Yamazaki, Kashu; Vo-Ho, Viet-Khoa; Bulsara, Darshan; Le, Ngan (July 2022). "Spiking Neural Networks and Their Applications: A Review". Brain Sciences. 12 (7): 863. doi:10.3390/brainsci12070863. ISSN 2076-3425. PMC 9313413. PMID 35884670.

9. Gerstner W (2001). "Spiking Neurons". In Maass W, Bishop CM (eds.). Pulsed Neural Networks. MIT Press. ISBN 978-0-262-63221-8.

10. Gerstner W, Kistler WM (2002). Spiking neuron models: single neurons, populations, plasticity. Cambridge, U.K.: Cambridge University Press. ISBN 0-511-07817-X. OCLC 57417395.

11. Kosmidis EK, Contoyiannis YF, Papatheodoropoulos C, Diakonos FK. Traits of criticality inmembrane potential fluctuations of pyramidal neurons in the CA1 region of rat hippocampus. Eur J Neurosci. 2018;48:235–2343.
https://doi.org/10.1111/ejn. 14117.

12. Yiannis F. Contoyiannis ,Efstratios K. Kosmidis, Fotios K. Diakonos, Myron Kampitakis, Stelios M. Potirakis.  A

hybrid artificial neural network for the generation of critical fluctuations and inter -spike intervals..Chaos, Solitons and Fractals. 159. 2022.

13. Bennett MV (1966). "PHYSIOLOGY OF ELECTROTONIC JUNCTIONS*". Annals of the New York Academy of Sciences. 137 (2): 509–539. Bibcode:1966NYASA.137..509B. doi:10.1111/j.1749-6632.1966.tb50178.x. ISSN 0077-8923. PMID 5229812.

14. Kandel ER, ed. (2013). Principles of neural science (5th ed.). New York: McGraw-Hill. ISBN 978-0-07-139011-8

15. Purves D, Williams SM, eds. (2004). Neuroscience (3rd ed.). Sunderland, Mass: Sinauer Associates. ISBN 978-0-87893-725.

16. Alukdar, S; Emdad, L; Das, SK; Fisher, PB (2 January 2022). "GAP junctions: multifaceted regulators of neuronal differentiation". Tissue Barriers. 10 (1) 1982349. 8794256. PMID 34651545.

17. Bennett MV, Zukin R (2004). "Electrical Coupling and Neuronal Synchronization in the Mammalian Brain". Neuron. 41 (4): 495–511. doi:10.1016/s0896-6273(04)00043-1. ISSN 0896-6273. PMID 14980200

18. Makarenko V, Llinás R (1998-12-22). "Experimentally determined chaotic phase synchronization in a neuronal system". Proceedings of the National Academy of Sciences. 95 (26): 15747–15752. Bibcode:1998PNAS...9515747M. doi:10.1073/pnas.95.26.15747. ISSN 0027-8424. PMC 28115. PMID 9861041.

19. Korn H, Faure P (2003). "Is there chaos in the brain? II. Experimental evidence and related models". Comptes Rendus. Biologies (in French). 326 (9): 787–840. doi:10.1016/j.crvi.2003.09.011. ISSN 1768-3238. PMID 14694754.

31